\documentclass[aps,pra,preprint,showpacs,superscriptaddress,amsfonts,floatfix]{revtex4}
\usepackage{graphics}

\usepackage[dvips]{graphicx}
\usepackage{amsmath}

\newcommand{\Tr}{\mathop{\mathrm{Tr}}}

\begin{document}

%
%

\title{Correlated model atom in a time-dependent external field: \\
Sign effect in the energy shift}

%
%
%
\author{I. Nagy}
\affiliation{Department of Theoretical Physics,
Institute of Physics, \\
Budapest University of Technology and Economics, \\ H-1521 Budapest, Hungary}
%
\author{ I. Aldazabal}
\affiliation{Centro de F\'{i}sica de Materiales (CSIC-UPV/EHU)-MPC,
P. Manuel de Lardizabal 5, \\ E-20018 San Sebasti\'an, Spain}
%

\date{\today}
\begin{abstract}

In this contribution we determine the exact solution for the ground-state wave function of a
two-particle correlated model atom with harmonic interactions. From that wave function, the
nonidempotent one-particle reduced density matrix is deduced. Its diagonal gives the exact
probability density, the basic variable of Density-Functional Theory. The one-matrix is
directly decomposed, in a point-wise manner, in terms of natural orbitals and their occupation numbers,
i.e., in terms of its eigenvalues and normalized eigenfunctions. The exact informations are used
to fix three, approximate, independent-particle models. Next, a time-dependent external
field of finite duration is addded to the exact and approximate Hamiltonians and the
resulting Cauchy problem is solved. The impact of the external field is investigated by calculating
the energy shift generated by that time-dependent field. It is found that
the nonperturbative energy shift reflects the sign of the driving field. 
The exact probability density and current are used, as inputs, to investigate the capability
of a formally exact independent-particle modeling in time-dependent DFT as well.
The results for the observable energy shift are analyzed
by using realistic estimations for the parameters of the two-particle target and the external field.
A comparison with the experimental prediction on the sign-dependent energy loss of swift protons and
antiprotons in a gaseous He target is made.

\end{abstract}

\centerline{\it Dedicated to the memory of Professor Rufus Ritchie}


\maketitle

\section{Introduction}

The determination of the energy lost by a fast charged particle by interacting with target
atoms through or near which it is passing has been the object of continuous interest. 
The essential featuress of the phenomenon are explained, before quantum mechanics, in terms
of a classical theory due to Bohr \cite{Bohr13}. He treated the electrons near which the particle
passes as classical oscillators that are set in motion by the (dipole) electric field of the
heavy passing particle. The energy thus absorbed by the electrons is equal to the energy
lost by the projectile. Besides this distant collision regime, close collisions were treated
as pure Rutherford scattering by neglecting electron binding. However, as Fermi pointed out,
quantum mechanical corrections have to be introduced for the very close
impact, when the fast particle passes {\it through} a target atom \cite{Fermi40}. 
Close to that regime, the system of
a heavy negative particle and a hydrogen atom has a critical distance
at which the binding energy of the electron becomes zero.  Its value is $0.639$ Bohr radii
according to Fermi and Teller \cite{Teller47}. They considered the capture of 
negative mesotrons, i.e.,  negative muons ($\mu^{-}$), in matter.

Range measurements with pions ($\pi^{\pm}$),
performed at around 1 MeV/amu in nuclear emulsions, predicted that the positive meson stopping power
is larger than the negative pion stopping power \cite{Barkas63}. Such measurements motivated
the pioneering theoretical work of the group of Ritchie \cite{Ritchie72} on the {\it charge-sign-effect}
in stopping. We note that experiments, performed with 
protons ($Z_1=1$) and antiprotons ($Z_1=-1$) at CERN, have confirmed this charge-sign-effect 
in several solid-state targets over a large range of impact energies \cite{Morenzoni89,Moller97}.

Ritchie's pioneering paper uses the harmonic oscillator as a model for an active electron bound
in an atom and deals only with distant collision considering the dipole and quadrupole terms
of the charged-projectile field. They considered their calculation as 
the classical equivalent of the second-order Born approximation. 
The subsequent, detailed quantum-mechanical calculation of Merzbacher \cite{Merzbacher74} 
on the same system results in agreement with that classical calculation.
In one, quantum mechanical, method the dipole interaction is taken into account
using a forced harmonic oscillator on which the quadrupole interaction is treated
as a first-order perturbation. However, concerning the relative importance of distant and
close collisions in a charge-sign effect, these calculations \cite{Ritchie72,Merzbacher74}
prescribe different values for the minimum impact parameter
below which they applied, without {\it binding-effect}, pure 
three-dimensional Rutherford scattering as in Bohr's early treatment.  

Motivated by these important works, in this paper, dedicated to the memory of Ritchie,
we will discuss a problem beyond the common (mean-field) one-electron treatment. We
would like to give a partial answer on the challenging question \cite{Quinteros91} of the interplay
of an inseparable interparticle correlation and a time-dependent
external excitation. In our exact treatment we will concentrate on
the sign-effect in the energy shift. This attempt is highly motivated by the
pioneering work of Ritchie, who used a single-electron modeling. We believe that
our estimations for close-impact could have relevance not only in physics, but, maybe more
importantly, in human cancer therapy \cite{Hori13} with antiproton beams as well. There
the precise knowledge of the biological effectiveness is vital. Clearly, close collisions
with {\it binding-effect} \cite{Kabachnik90,Grande97}, but without 
the annihilation process, could be important.

Parallel to this exciting atomistic challenge, we investigate the important problem
of a harmonically confined interacting bosonic system \cite{Demler12,Islam15}, where
the time-dependent confinement-tuning is an easily realizable experimental tool
to generate correlated dynamics. Such confinement-tuning can be continuous in
contrast to the discrete case with scattering of charged particles off an atom. In that,
actively investigated research field of bosons the information content, like the entropies
and cloud-overlaps, are in the focus of studies.

This paper is organized as follows. The next two Sections contain our
target-model and the formulation of the time-dependent problem. The results, with
discusions, are given in Section IV. The last Section is devoted to a short summary.
We will use atomic units.


\section{The target: A two-particle correlated model system}

As our unperturbed target, we take the interacting two-particle model 
first introduced by Heisenberg \cite{Heisenberg26} as one of the 
really simplest many-body models, and write
\begin{equation}
\hat{H}_0(x_1,x_2)\, =\, -\, \frac{1}{2}\left(\frac{d^2}{dx_1^2}\, +
\frac{d^2}{dx_2^2}\right) +\frac{1}{2}\, \omega_0^2({x}_1^2+{x}_2^2)
-\frac{1}{2}\, \lambda\, \omega_0^2({x}_1-{x}_2)^2,
\end{equation}
to the stationary Schr\"odinger equation. In Heisenberg's classification of
this Hamiltonian: it is {\it Das denkbar einfachste Mehrk\"orperproblem}.
In Eq.(1) $\lambda\in{[0,0.5]}$ measures the strength of the allowed (see, below) repulsive 
interparticle interaction energy in terms of $\omega_0^2$. 
For $\lambda>0.5$, both interacting particles 
cannot both remain in the confining external field.
Our exact treatment using a simple model atom will allow a useful diagnosis \cite{Cohen08}
of sophisticated one-electron approximations as well. These may suffer from failures in prediction.

Introducing standard \cite{Moshinsky68,Davidson76,Pipek09} normal 
coordinates $X_1\equiv{(x_1+x_2)/\sqrt{2}}$ 
and $X_2\equiv{(x_1-x_2)/\sqrt{2}}$,
one can easily rewrite the unperturbed Hamiltonian into the form 
\begin{equation}
\hat{H}_0(X_1,X_2)\, =\, -\, \frac{1}{2}\left(\frac{d^2}{dX_1^2}\, +
\frac{d^2}{dX_2^2}\right) +\frac{1}{2}\, \omega_1^2\, X_1^2 + \frac{1}{2}\, \omega_2^2\, X_2^2,
\end{equation}
where  $\omega_1\equiv{\omega_0}$ and $\omega_2\equiv{\omega_0\sqrt{1-2\lambda}}$ denote the 
frequencies of the independent normal modes. 
Based on Eq.(2), the normalized ground-state wave function $\Psi(X_1,X_2)$ is a product
\begin{equation}
\Psi(X_1,X_2) = \phi_1(X_1)\phi_2(X_2)\equiv{
\left(\frac{\omega_1}{\pi}\right)^{1/4}\exp\left[-\frac{1}{2}\omega_1\, X_{1}^2\right]\, 
\left(\frac{\omega_2}{\pi}\right)^{1/4}\exp\left[-\frac{1}{2}\omega_2\, X_{2}^2\right]}. 
\end{equation}
Notice that $\phi_1(X_1)\phi_2(X_2)\neq{\phi_1(x_1)\phi_2(x_2)}$.
We stress that the price of this normal-mode transformation is that one loses
the intuitive physical picture
of {\it real} particles and, instead, operates with {\it effective} independent particles 
representing the transformed
coordinates. 
The ground-state energy of this model is $E=(1/2)(\omega_1+\omega_2)$. There is
an equal contribution from the kinetic [$(1/4)(\omega_1+\omega_2)$] and potential 
[$(1/4)(\omega_1+\omega_2)$] parts in accord with the virial theorem for bounded states.
The first ionization energy is given by $\omega_2=\omega_0\sqrt{1-2\lambda}$. 
A formal change, $X_1\rightarrow{x_1}$ and $X_2\rightarrow{x_2}$ in Eq.(3) would
correspond to a two-active-particle modeling, where, in an independent-particle picture,
the ionization energies are pre-fixed.

Working with approximations at the wave-function level, we can take a product
\begin{equation}
\Psi_e(x_1,x_2)\, =\, \left(\frac{\omega_e}{\pi}\right)^{1/4}\exp\left[-\frac{1}{2}\omega_e\, x_{1}^2\right]\, 
\left(\frac{\omega_e}{\pi}\right)^{1/4}\exp\left[-\frac{1}{2}\omega_e\, x_{2}^2\right]. 
\end{equation}
as a parametric ($\omega_e$) state and perform total-energy ($e$) optimization with the 
original Hamiltonian in Eq.(1).
In fact, such an attempt is the Hartree-Fock approximation \cite{Moshinsky68,Davidson76} 
and we get \cite{Pipek09} after minimization 
$\omega_e=\omega_0\sqrt{1-\lambda}$ to Eq.(4). 
The corresponding, i.e., energetically optimal, independent-particle 
Hamiltonian behind $\Psi_e(x_1,x_2)\equiv{\phi_e(x_1)\phi_e(x_2)}$ is
\begin{equation}
\hat{H}_0^{(e)}(x_1,x_2)\, =\, -\, \frac{1}{2}\left(\frac{d^2}{dx_1^2}\, +
\frac{d^2}{dx_2^2}\right) +\frac{1}{2}\, \omega_e^2({x}_1^2+{x}_2^2).
\end{equation}

By rewriting the wave function $\Psi(X_1,X_2)$ in terms of physical coordinates $x_1$ and $x_2$, we
determine the statistical one-matrix $\Gamma_1(x_1,x_2)$ via the following 
\cite{Davidson76,Pipek09} nonlinear mapping
\begin{equation}
\Gamma_1(x_1,x_2)\, =\, \int_{-\infty}^{\infty}\, dx_3\,
\Psi^{*}(x_1,x_3)\, \Psi(x_2,x_3). \nonumber
\end{equation}
From this, we arrive at an informative, i.e., Jastrow-like, representation
\begin{equation}
\Gamma_1(x_1,x_2)=\phi_d(x_1)\, \phi_d(x_2) 
\times{e^{-\frac{1}{2}D[(x_1-x_2)]^2}}
\end{equation}
where, with $\omega_d\equiv{2\omega_1\omega_2/(\omega_1+\omega_2})$, we 
introduced the following abbreviations
\begin{equation}
\phi_d(x)=\left[\frac{\omega_d}{\pi}\right]^{1/4}\,
e^{-\frac{1}{2}\omega_d\, x^2}  \nonumber   
\end{equation}
\begin{equation}
D\, =\, \frac{1}{4}\, \frac{(\omega_1-\omega_2)^2}
{\omega_1+\omega_2}\, \geq{0}. \nonumber
\end{equation}
The diagonal ($x_1=x_2=x$) of the nonidempotent $\Gamma_1(x_1,x_2)$ gives the one-particle 
probability density, $n(x)=\Gamma_1(x,x)$, of unit-norm. 
Thus, in the knowledge of this exact result for the density $n(x)$,
one may introduce the second, density-optimal ($d$), auxiliary product
\begin{equation}
\Psi_d(x_1,x_2)\, =\, \phi_d(x_1)\phi_d(x_2)\, = \, \phi_d(X_1)\phi_d(X_2),
\end{equation}
and associate with it an independent-particle Hamiltonian via inversion
\begin{equation}
\hat{H}_{0}^{(d)}(x_1,x_2)\, =\, -\, \frac{1}{2}\left(\frac{d^2}{dx_1^2}\, +
\frac{d^2}{dx_2^2}\right) +\frac{1}{2}\, \omega_d^2({x}_1^2+{x}_2^2).
\end{equation}
This Hamiltonian contains, upto an undetermined \cite{Kosugi18} constant $C_1(\lambda)$, the 
Kohn-Sham potential in the orbital implementation of Density Functional Theory \cite{Dreizler90}.
One may consider such an implementation as a first-principle method based on semi-empirical
inputs \cite{Cohen08}. In the knowledge of the exact ground-state energy, one could fix this
constant in an inversion as $C_1(\lambda)\equiv{(1/4)(\omega_0+\omega_2)-(1/2)\omega_d}=
(1/4)(\omega_0-\omega_2)^2/(\omega_0+\omega_2)$, thus $C_1(\lambda=0)=0$.

Consider, finally, the Jastrow-like form in Eq.(6). We can apply a point-wise \cite{Riesz55} 
direct decomposition for it. Indeed, Mehler's formula \cite{Erdelyi53,Robinson77} reads
\begin{equation}
(\omega_w/\pi)^{1/2}\,  e^{- \frac{\omega_w}{2}  
\left(\frac{1+Z^2}{1-Z^2}\right) (x_1^2+x_2^2)}\,  e^{\omega_w \frac{2\, Z}{1-Z^2}\, x_1 x_2}=
\sum_{k=0}^{\infty} (1-Z^2)^{1/2}\, Z^k\, \phi_k(\omega_w,x_1)\, \phi_k(\omega_w,x_2),
\end{equation}
where the parameter $Z\in{[0,1]}$, and $x_i\in{(-\infty,\infty)}$. The $\phi_k(\omega_w,x)$ 
decomposition-functions form a 
complete set of orthonormal eigenfunctions of a one-dimensional harmonic oscillator 
with potential energy 
$\omega_w^2(x^2/2)$ in the Schr\"odinger wave equation and are given by
\begin{equation}
\phi_k(\omega_w,x)\, = \,
\left(\frac{\omega_w}{\pi}\right)^{1/4}\frac{1}{\sqrt{2^k\,
k!}}\, e^{-\frac{1}{2}\omega_w\, x^2}\,
H_k(\sqrt{\omega_w}x).
\end{equation}
Comparison of exponentials in Eq.(6) and Eq.(9) results in the two constraints 
\begin{equation}
(\omega_d+D)\, =\, \omega_w\, \frac{1+Z^2}{1-Z^2} \nonumber
\end{equation}
\begin{equation}
D\, =\, \omega_w\, \frac{2Z}{1-Z^2}. \nonumber
\end{equation}
One can solve these equations easily for $Z$ and $\omega_w$ in terms 
of $D$ and $\omega_d$. We get
\begin{equation}
Z(\lambda)\, =\, \frac{\sqrt{1+2D/\omega_d}-1}{\sqrt{1+2D/\omega_d}+1}\, =\, 
\left(\frac{\sqrt{\omega_1}-\sqrt{\omega_2}}{\sqrt{\omega_1}+\sqrt{\omega_2}}\right)^2 \nonumber
\end{equation}
\begin{equation}
\omega_w\, =\, \omega_d\sqrt{1+2D/\omega_d}\, \equiv{\sqrt{\omega_1\omega_2}}
=\omega_0(1-2\lambda)^{1/4}. \nonumber
\end{equation}

Since $\omega_d=\omega_w(1-Z)/(1+Z)$, we obtain a closed-shell-like expansion
\begin{equation}
\Gamma_1(x_1,x_2,\omega_w)\, =\, \sum_{k=0}^{\infty}\, P_k(Z)\,
\phi_k(\omega_w,x_1)\, \phi_k(\omega_w,x_2),
\end{equation}
where the occupation numbers of the {\it natural} \cite{Lowdin56} orbitals, 
$\phi_k(\omega_w,x)$, are 
\begin{equation}
P_k(Z)\, =\, (\omega_d/\omega_w)^{1/2}\, (1-Z^2)^{1/2} Z^k\, =\, (1-Z) Z^k. \nonumber
\end{equation}
Of course, we have $\sum_{k=0}^{\infty}P_k=1$. The exact result in Eq.(11) suggests our 
third, so-called wave-function-optimal ($w$), independent-particle modeling with
$\phi_w(x)\equiv{\phi_0(\omega_w,x)}$ in
\begin{equation}
\Psi_w(x_1,x_2)\, =\, \phi_w(x_1)\phi_w(x_2).
\end{equation}
\begin{equation}
\hat{H}_{0}^{(w)}(x_1,x_2)\, =\, -\, \frac{1}{2}\left(\frac{d^2}{dx_1^2}\, +
\frac{d^2}{dx_2^2}\right) +\frac{1}{2}\, \omega_w^2({x}_1^2+{x}_2^2).  
\end{equation}
With the above product for $\Psi_w(x_1,x_2)$ the overlap with the exact $\Psi(x_1,x_2)$ 
is maximal variationally \cite{Nagy11}. At $\lambda\neq{0}$ we have 
the following ($\omega_1\equiv{\omega_0}$) ordering of frequencies: 
\begin{equation}
\omega_2<\omega_d<\omega_w<\omega_e<\omega_1.
\end{equation}
The ordering $\omega_d<\omega_w<\omega_e$ shows that the density-optimal ($d$) Kohn-Sham method 
and the energy optimal ($e$) Hartree-Fock method under- and overestimate the localization \cite{Cohen08}.
Besides, there is an unphysical saturation in $\omega_e$ at the physically allowed
$\lambda\rightarrow{0.5}$ limit.

Notice that one may use our normalized distribution $P_k$ to calculate various
information-theoretic entropies (R\'enyi and von Neumann) or to define a normalized 
escort distribution as input to nonextensive (Tsallis)
statistics via $P_k(Z)\rightarrow{P_k^{(q)}\equiv{P_k(Z^q)}}$. 
Entropies measure, in our case with an interacting Hamiltonian, the deviations from independent-particle
modelings, i.e., they reflect the inseparability of this Hamiltonian in original particle coordinates.
They have an inherent connection with the important spectral aspect of interparticle correlation.
Measuring entanglement characteristics is a recent hot topic of
fundamental theoretical and practical interest in 
laser-field-confined cold-atom systems \cite{Demler12,Islam15}.

\newpage

\section{The perturbation: a time-dependent quadrupolar field}

In quantum mechanics, in the presence of an external time-dependent perturbation 
of finite duration, one defines the {\it time-independent} energy shift from an
expectation value of the unperturbed Hamiltonian with the time-evolving state. This
state is the solution of the time-dependent Schr\"odinger equation where one has
a Cauchy problem with a prescribed initial state. In this path, the correct \cite{Ballentine98}
interpretation of the effect of a time-dependent peturbation on the target-system is to produce a
nonstationary state rather than to cause a jump from one stationary state to another.
However, following Dirac's method (variation of constants) to time-dependent problems, 
in an energy-shift calculation one can equally-well 
apply transition probabilities to the allowed excited states as occupation
probabilities (statistical weights)
to energetics. In this paper we will outline both interpretations,
and demonstrate their equivalence in the determination of a {\it time-independent} energy change.

To our Hamiltonians in Eq.(1), Eq.(5), Eq.(8), and Eq.(13) we add, as excitation 
of $\hat{H}'(t\rightarrow{-\infty})=\hat{H}'(t\rightarrow{\infty})=0$ 
character, a time-dependent external perturbation
\begin{equation}
\hat{H}'(x_1,x_2,t)\, =\, \frac{1}{2}\, \omega_0^2\, \Lambda\, F(t)\, ({x}_1^2+{x}_2^2)\equiv{
\frac{1}{2}\, \omega_0^2\, \Lambda\, F(t)\, (X_1^2+X_2^2)}
\end{equation}
where $\Lambda$ can be positive or negative in oder to discuss the sign-effect in energy shift.
Notice that in modern experiments on harmonically confined systems, with 
precisely specified number of atoms \cite{Islam15}, our peturbation could be considered as
confinement-tuning in time. The sum of the Hamiltonians in Eq.(1) and Eq.(15) defines,
mathematically, a so-called isospectral deformation of the ground-state Hamiltonian.

Since the exact Eq.(2) for two independent {\it modes} and the effective Hamiltonians in Eq.(5), Eq.(8), and
Eq.(13) for two independent {\it particles} have a separable behavior in $X_1$ and $X_2$ coordinates,
the addition of our separable $\hat{H}'(X_1,X_2,t)$ allows simplification in mathematics.
Clearly, one can consider a single oscillator [$\hat{h}(X,t)$] for which
\begin{equation}
\hat{h}(X,t)\, =\, -\, \frac{1}{2}\, \frac{d^2}{dX^2}\, + \frac{1}{2}\, \Omega^2(t)\, X^2 .
\end{equation}
where $\Omega^2(t)\equiv{[\Omega_0^2 +\omega_0^2\, \Lambda\, F(t)]}$,
and $\Omega_0$ is a shorthand for $\omega_1, \omega_2$ and $\omega_e,\omega_d,\omega_w$.
Thus, one has to solve the corresponding time-dependent Schr\"odinger equation
\begin{equation}
i\, \frac{\partial \phi(X,t)}{\partial t}\, =\, \hat{h}(X,t)\phi(X,t),
\end{equation}
considering the initial condition for $\phi(X)$ at $t=-\infty$ where $\hat{h}'(X,t)$ is zero.
Our problem is one of the rare cases where it is possible to solve a nonstationary problem
exactly. This mathematical exactness could be useful considering the
physical statements based on it. Taking established papers \cite{Popov70,Kagan96}
on the solution of Eqs.(16-17), we proceed along them and, similarly to these papers,
we exclude the $[\Omega_0^2 +\omega_0^2\, \Lambda\, F(t)]<0$ case, which may happen
with certain negative $\Lambda$. Thus, in the present work, where we will 
use $F(t=0)=1$ as maximum below, 
we restrict ourselves to $|\Lambda|<(\omega_2/\omega_0)^2=(1-2\lambda)$. An equality
would mimic, at negative $\Lambda$, an ionization-like situation at $t=0$.


\section{Results and discussion} 

The solution rests on making proper changes of the time and distance
scales \cite{Popov70,Kagan96} to consider time-evaluation in confinement frequency.
The nonstationary evolving state, denoted by $\phi[X,\Omega_0,B,t]$,
contains these scales as
\begin{equation}
\phi(X,\Omega_0,B,t)=\left[\frac{\Omega_0}{B^2(t)\, \pi}\right]^{1/4}\, \exp\left[-\frac{X^2}{2}\, 
\frac{\Omega_0}{B^2(t)}
\left(1-i\, \frac{B(t)\dot{B}(t)}{\Omega_0}\right)\right]\, e^{-i\, \gamma(t)/2},
\end{equation}
where $B(\Omega_0,t)$ and $\gamma(t)$ are interrelated in the complex solution, $\xi(t)=B(t)
\exp[i\gamma(t)]$, of
the following classical equation of motion
\begin{equation}
\ddot{\xi}(t) +\, \Omega^2(t)\, \xi(t)\, =\, 0,
\end{equation}
The nonlinear, Ermakov-type \cite{Ermakov08,Pinney50} differential equation, determining the real 
scale-function $B(t)=|\xi(t)|$ becomes
\begin{equation}
\ddot{B}(t) + B(t)\, \Omega^2(t)\, =\, B(t)[\dot{\gamma}(t)]^2\, =\, \frac{\Omega_0^2}{[B(t)]^3},
\end{equation}
after taking $\dot{\gamma}(t)\equiv{\Omega_0/B^2(t)}$. The two initial
conditions, reflecting the behavior of our passing excitation, are 
$B(t=-\infty)=1$ and $\dot{B}(t=-\infty)=0$. Notice that one could linearize this
nonlinear equation in four steps by taking first $A(t)\equiv{B^2(t)}\neq{0}$. 
However, the resulting linear differential equation
\begin{equation}
\dddot{A}(t)+4\Omega^2(t)\dot{A}(t)+4\Omega(t)\dot{\Omega}(t)A(t)\, =\, 0, \nonumber
\end{equation}
becomes a third-order one which would need three initial conditions.

\newpage

Now, we illustrate the entangled nature of our correlated two-particle system in the time domain
as well, similarly to the stationary case in Section II. 
By rewriting the exact wave function $\Psi(X_1,X_2,t)=\phi_1[X_1,\omega_0,B(\omega_0,t),t]
\phi_2[X_2,\omega_2,B(\omega_2,t),t]$
in terms of original coordinates, we
determine the reduced single-particle density matrix from
\begin{equation}
\Gamma_1(x_1,x_2,t)\, =\, \int_{-\infty}^{\infty}\, dx_3\, \nonumber
\Psi^{*}(x_1,x_3,t)\, \Psi(x_2,x_3,t),
\end{equation}
After a long, but straightforward calculation we obtain
\begin{equation}
\Gamma_1(x_1,x_2,t)=\phi_d(x_1,t)\, \phi_d^{*}(x_2,t)\, \times{
e^{-\frac{1}{2}D(t)(x_1-x_2)^2}}, \nonumber
\end{equation}
where we introduced, as a generalization of the static case, the following abbreviations
\begin{equation}
\phi_d(x,t)=\left[\frac{\omega_d(t)}{\pi}\right]^{1/4}\,
e^{-\frac{1}{2}\omega_d(t)\, x^2[1-i\, \alpha(t)/\omega_d(t)]}, 
\end{equation}
\begin{equation}
D(t)\, =\, \frac{1}{4}\, \frac{[\omega_1(t)-\omega_2(t)]^2
+[\dot{B}(\omega_1,t)/B(\omega_1,t) -
\dot{B}(\omega_2,t)/B(\omega_2,t)]^2} {\omega_1(t)+\omega_2(t)} \geq{0} \nonumber
\end{equation}
\begin{equation}
\frac{\alpha(t)}{\omega_d(t)}\, =\, \frac{1}{2}\,
\left[\frac{B(\omega_1,t)\dot{B}(\omega_1,t)}{\omega_1}+
\frac{B(\omega_2,t)\dot{B}(\omega_2,t)}{\omega_2}
\right] \nonumber
\end{equation}
with $\omega_d(t)=2\omega_1(t)\omega_2(t)/[\omega_1(t)+\omega_2(t)]$,
where $\omega_i(t)=\omega_i/[B(\omega_i,t)]^2$ and $i=1,2$. There is
{\it mode-mixing} in $\omega_d(t)$, $D(t)$, and $\alpha(t)$.
Following our evaluation at Eqs.(6-11) in the stationary case, here
we add the final form for the time-dependent occupation numbers
\begin{equation}
P_k[Z(t)]\, =\, [1-Z(t)]\, [Z(t)]^k, \nonumber
\end{equation}
\begin{equation}
Z(t)\, =\, \frac{\sqrt{1+2D(t)/\omega_d(t)}-1}{\sqrt{1+2D(t)/\omega_d(t)}+1}, \nonumber
\end{equation}
in terms of the time-dependent variables $D(t)$, and $\omega_d(t)$. This
normalized distribution function could allow an analysis of time-dependent
information-theoretic entropies \cite{Nagy18}.

Our nonidempotent one-matrix, with operator-trace $\Tr[\Gamma^2_1(t)]=[1+2D(t)/\omega_d(t)]^{-1/2}<1$,
is Hermitian since $\Gamma_1(x_1,x_2,t)=\Gamma_1^{*}(x_2,x_1,t)$. 
Its diagonal, where $x_1=x_2=x$, gives the exact one-particle 
probability density $n(x,t)=\Gamma_1(x,x,t)$, i.e., the basic variable \cite{Dreizler86} 
of time-dependent DFT.
The exact probability current becomes $j(x,t)=x\, n(x,t)\, \alpha(t)$ and, of course, the 
continuity equation of quantum mechanics is satisfied $\partial_t n(x,t)+\partial_x j(x,t)=0$.
We strongly stress at this point, that
in $\phi_d(x,t)$ , i.e., in a density-  and current-optimal Kohn-Sham-like auxiliary orbital 
we have
$\omega_d(t)\neq{\omega_d/[B(\omega_d,t]^2}\equiv{\bar{\omega}_d(t)}$ at $t\neq{-\infty}$.

\newpage

These exact probabilities are needed, in the two-particle case, to invert \cite{Lein05} a Schr\"odinger-like
time-dependent Kohn-Sham equation for its doubly-populated orbital in order to find 
an effective time-dependent potential behind that auxiliary orbital $\phi_d(x,t)$, 
at least upto an undetermined time-dependent constant $C_2(\lambda,t)$. From such an inversion
we obtain
\begin{equation}
V_d(x,t) = \frac{1}{2\sqrt{n(x,t)}}\, \partial_x^2 \sqrt{n(x,t)} -\frac{1}{2}[\partial_x k(x,t)]^2
-\partial_t k(x,t) \nonumber
\end{equation}
where $k(x,t)\equiv{(1/2) x^2 \alpha(t)}$ to the polar form 
$\phi_d(x,t)=\sqrt{n(x,t)}\exp[ik(x,t)]$. The first term on the right-hand-side is an adiabatic
potential which produces $n(x,t)$ as its instantaneous ground-state density. 
In our two-particle model with harmonic interactions, we obtain, as with
Eqs.(6.50-6.52) on page 105 of \cite{Ullrich12}, 
for the effective potential after substitution
\begin{equation}
V_d(x,t) = \frac{1}{2}\, x^2\, \omega_d^2(t)\, -\frac{1}{2}\, 
x^2\sqrt{\omega_d(t)}\, \frac{d^2}{dt^2}\left[\frac{1}{\sqrt{\omega_d(t)}}\right].
\end{equation}
This, {\it formally} exact, effective potential may depend \cite{Maitra15} 
on time even at $t\rightarrow{\infty}$, 
which is unphysical [see, Eq.(25) below]
in the case of a perturbation with finite duration. 
However, starting from a pre-optimized
state given by Eq.(7), and using in the above form the corresponding time-dependent frequency
$\bar{\omega}_d(t)\equiv{\omega_d/[B^2(\omega_d,t)]}$ instead of $\omega_d(t)$, we get
\begin{equation}
\bar{V}_d(x,t) = \frac{1}{2}\, x^2\, 
\left[\frac{\omega^2_d}{B^4(\omega_d,t)} -\frac{\ddot{B}(\omega_d,t)}{B(\omega_d,t)}\right]\,
\equiv{\frac{1}{2}\, x^2\, [\omega_d^2+\Lambda \omega_0^2 F(t)]}, \nonumber
\end{equation}
because of Eq.(20). Clearly, this potential behaves as it should {\it physically}.


Our main goal in this study is the determination of the time-independent energy shift 
generated by a time-dependent excitation
switched on and off, i.e., which has a symmetric
$\hat{H}'(t\rightarrow{-\infty})=\hat{H}'(t\rightarrow{\infty})=0$ character. To this determination
we need the solution for $B(t)$ at the {\it long-time} limit where the the frequency is already $\Omega_0$. 
By considering the fact 
that Eq.(19), with $\xi(t\rightarrow{-\infty})=e^{i\Omega_0 t}$, is mathematically
equivalent \cite{Popov70,Kagan96} to a 
Schr\"odinger equation for the reflection coefficient $R(\Omega_0)$ of a particle with "energy" 
$\Omega_0^2$ in the "potential" energy
$-\Lambda\omega_0^2 F(t)$, if $t$ is interpreted as a "spatial coordinate", one can \cite{Kagan96} write
\begin{equation}
B^2(\Omega_0,t)=|\xi(t)|^2=\frac{1+R}{1-R}\, -\, \frac{2\sqrt{R}}{1-R}\, \cos(2\Omega_0 t+\delta),
\end{equation}
\begin{equation}
\gamma(t)\, =\, \arctan\left[\frac{1+\sqrt{R}}{1-\sqrt{R}}\tan(\Omega_0 t+\delta/2)\right]. \nonumber
\end{equation}
Of course, we have $\dot{\gamma}(t)=\Omega_0/B^2(\Omega_0,t)$. The
form in Eq.(23) is exact at $t\rightarrow{\infty}$.
Clearly, once an $F(t)$ is given and the corresponding $R$ is found, 
we can go back to the time-evolving wave function $\phi(X,t)$
and use it to calculate energy expectation value with the time-independent Hamiltonian $\hat{h}(X)$. 
After subtraction from that expectation value the ground-state energy $\Omega_0/2$, 
one can find the time-independent energy shift $\Delta E(\Omega_0)$.

It is easy to show, by applying Eq.(23) to the evolving solution in Eq.(18), that only 
the total energy becomes time-independent. Its kinetic and potential contributions
\begin{equation}
<K(t)>\, \, =\, \, \frac{\Omega_0}{4}\,
\left[\frac{1}{B^2(\Omega_0,t)}\, +
\frac{\dot{B}^2(\Omega_0,t)}{\Omega_0^2} \right] \nonumber
\end{equation}
\begin{equation}
<V(t)>\, \, =\, \frac{\Omega_0}{4}\,
B^2(\Omega_0,t). \nonumber
\end{equation}
are oscillating functions at the physically important $t\rightarrow{\infty}$ long-time limit. 
The one-mode, time-independent energy change takes, by using Eq.(23), a remarkably simple form
\begin{equation}
\Delta E(\Omega_0)\, \equiv{ \left[\frac{1}{2}\, \Omega_0\, \frac{1+R}{1-R} - 
\frac{1}{2} \, \Omega_0\right]}\, = \, \Omega_0\, \frac{R}{1-R}.
\end{equation}
We stress that although $B(\Omega_0,t)$ is an oscillating function, to energy shift we have
\begin{equation}
\frac{1}{4}\, \frac{1}{B^2(t)} \left[1 +\left(\frac{B(t)\dot{B}(t)}{\Omega_0}\right)^2 + B^4(t)\right]\, 
=\, \frac{1}{2}\, \frac{1+R(\Omega_0)}{1-R(\Omega_0)}. \nonumber
\end{equation}
Only this combination of $B(t)$ and $\dot{B}(t)$ will be
independent of the time. Notice that using the corresponding time-independent mean values 
[$\bar{B^4}\neq{(\bar{B^2})^2}$] 
of our oscillating functions in the above expression, we recover the exact independent-mode result since
\begin{equation}
\frac{1}{4}\, \frac{1-R}{1+R}\, \left[1 + \frac{2R}{(1-R)^2} + \frac{1+4R+R^2}{(1-R)^2}\right]\,=\,
\frac{1}{2}\, \frac{1+R}{1-R} \nonumber
\end{equation}

When we calculate the {\it total} energy of our two-particle system, with Eq.(21) for
the density-optimal wave function and Eq.(22) for the corresponding effective
potential, as an expectation value in quantum mechanics, we arrive at the following
\begin{equation}
<E_d(t)> \, =\, \frac{1}{2}\, \omega_d(t)\left[1+\frac{\alpha^2(t)}{\omega_d^2(t)}\right]
+ \frac{1}{2}\, \omega_d(t)\left[1-\frac{1}{\omega_d^{3/2}(t)}\frac{d^2}{dt^2}
\left(\frac{1}{\sqrt{\omega_d(t)}}\right)\right],
\end{equation}
where the first expectation value is the kinetic part, 
and second expectation value is the potential part.
This total energy is not time-independent at $t\rightarrow{\infty}$. 
It oscillates (at $\lambda\neq{0}$) around its mean value 
reflecting \cite{Maitra15} the oscillating behavior of the basic variable $n(x,t)$. One might
consider this mean value (an average, based on mixed modes) as a steady-state value.
But, the such-defined time-independent energy shift
is {\it not} exact. Its deviation from the exact 
energy shift measures, {\it a posteriori}, the quality of approximations in TDDFT.

We note at this subtle point that one may argue that the $C_2(\lambda,t)$ 
constant [not determined via inversion, but 
$C_2(\lambda,t=-\infty)=C_1(\lambda)]$ could cure such a failure. Unfortunately,
we have no rule to construct that time-dependent constant {\it a priori}, i.e., without
the knowledge of the exact answer for the time-independent total energy.
In complete agreement
with Dreizler's early remark \cite{Dreizler86} made at around the foundation of TDDFT, there
are several reasons why the {\it concept} of universal functionals for time-dependent
systems will play an even less important role in practice than
in the time-independent case. We will return to this point at Eq.(35), where we calculate
an action-like quantity with our exact wave function.

Now we outline briefly the second method, discussed at the beginning of Section III, which
is based on transition probabilities. Interpreting these as occupation numbers one can perform
a statistical averaging. These statistical weights are given by
\begin{equation}
W_{2n,0}(R)\, =\, \frac{(2n)!}{2^{2n}(n!)^2}\, \sqrt{1-R}\, (R)^n = 
\frac{1}{\sqrt{\pi}}\frac{\Gamma(n+1/2)}{\Gamma(n+1)}\sqrt{1-R}\, (R)^n, 
\end{equation}
considering the allowed (by a selection rule) transitions from the ground-state \cite{Popov70}. This
distribution function is normalized, i.e., $\sum_{n=0}^{\infty}W_{2n,0}(R)=1$, since
\begin{equation}
\frac{1}{(1-x)^{\eta}}\, =\, \sum_{n=0}^{\infty}\, 
\frac{\Gamma(n+\eta)}{n! \Gamma(\eta)}\, x^n. \nonumber
\end{equation}
By using this and the identity $n x^n=x (x^n)'$, the
energy shift can be easily calculated as a properly weighted sum in an upward process,
and the result becomes
\begin{equation}
\Delta E(\Omega_0)\, =\, \Omega_0\, \left[\sum_{n=0}^{\infty}\, 
\left(2n+\frac{1}{2}\right)W_{2n,0}[R(\Omega_0)] -\frac{1}{2}\right]\, \equiv{\Omega_0\, \frac{R(\Omega_0)}
{1-R(\Omega_0)}}.
\end{equation}
The equivalence of the two interpretations behind an excitation process is demonstrated.

In order to implement our exact framework, we apply an analytically solvable 
\cite{Landau58,Peierls79} textbook modeling for the time-dependence in the external field
\begin{equation}
F(t)\, =\, \frac{1}{\cosh^2(2\beta t)}, 
\end{equation}
and calculate $R(\Omega_0)$ of the associated "scattering" problem discussed at Eq.(23). 
One can consider $(1/\beta)$ as an effective transition time. In fast, charged-particle
penetration (with velocity $v$) through atomistic targets
the $\beta\propto{v}$ choice seems to be a reasonable one \cite{Artacho07}.
The one-mode, time-independent energy shift in Eq.(24) becomes
\begin{equation}
\Delta E(\Omega_0)\, = \, \Omega_0\, \frac{R}{1-R}\, =\,  \frac{ 
\Omega_0} {\sinh^2[(\pi/2)(\Omega_0/\beta)]}\, \cos^2\left[(\pi/2)
\sqrt{1+\Lambda\omega_0^2/\beta^2}\right]. 
\end{equation}
When $\Lambda \omega_0^2/\beta^2<-1$, which may happen for negative $\Lambda$, one must take
$\cos(i u)\Rightarrow{\cosh(u)}$.
In agreement with general rules of time-dependent perturbation theory, the energy shift tends 
to zero in the adiabatic ($\beta\rightarrow{0}$) and 
sudden ($\beta\rightarrow{\infty}$) limits. 
This closed expression  
shows that the energy shift can be an {\it oscillating} function of $\Lambda>0$ at fixed
$\beta$. The zeros are at $(2n+1)^2=[1+\Lambda(\omega_0/\beta)^2]$, where $n$ is a nonzero integer.
$Z_1$-oscillation in the energy-loss, i.e., in energy deposition, as a function of the 
nuclear charge ($Z_1>1$) {\it and} velocity ($\beta\propto{v}$) of intruders moving 
in condensed matter is a challenging
question \cite{Vincent06,Hatcher08,Puska14} of nonperturbative character. In our present modeling
of the impact of a time-dependent perturbation the first-order, Born-like $(B)$ result corresponds
to the limit of $\Lambda\rightarrow{0}$ in Eq.(29), and it becomes
\begin{equation}
\Delta E^{(B)}(\Omega_0)\, = \, \left(\frac{\Lambda \omega_0^2\pi}{4\beta^2}\right)^2\, 
\frac{ \Omega_0} {\sinh^2[(\pi/2)(\Omega_0/\beta)]}. \nonumber 
\end{equation}
This perturbative result does not depend on the sign ($\Lambda$) of the external field.

In the fast-excitation, i.e., sudden limit, we get from Eq.(29) by a careful expansion
\begin{equation}
\Delta E(\Omega_0)\simeq{\Omega_0}\, \left(\frac{\Lambda \omega_0^2}{2}\right)^2\, 
\frac{1}{(\beta \Omega_0)^2}\,
\left[1-\frac{1}{3}\left(\frac{\pi\Omega_0}{2\beta}\right)^2\right]\,
\left(1-\frac{1}{\beta^2}\, \frac{\Lambda \omega_0^2}{2}\right).
\end{equation}
This asymptotic form  exhibits a sign effect, 
since $\Lambda$ can be positive or negative.  However, this sign effect ($\propto{-\Lambda \omega_0^2/v^2}$)
is {\it opposite}
to the Barkas effect ($\propto{Z_1\Omega_0^2/v^3}$) found earlier \cite{Ritchie72,Merzbacher74}
in the dipole limit with charge-conjugated particles $Z_1=\pm{1}$. In agreement with Fermi's forecast
\cite{Fermi40}, our theoretical result signals the importance of a quantum mechanical 
treatment for fast ''close collisions'' with consideration ($\omega_0\neq{0}$)
of the binding effect \cite{Kabachnik90,Grande97}.
Such, reversed Barkas effect was predicted earlier
using data sets from OBELIX experiments at CERN with swift ($v\simeq{5-6}$)
protons and antiprotons moving in gaseous He, i.e., in a target 
of compact inert atoms \cite{Rizzini04}.
We will return to this data-based prediction at Fig. 3, at the end of this section.
Due to the factorized form in Eq.(29), in our three independent-particle approximations 
with pre-determined product states in Eq.(4), Eq.(7), and Eq.(12) we get
\begin{equation}
\Delta E(\Omega_0=\omega_e)<\Delta E(\Omega_0=\omega_w)<\Delta E(\Omega_0=\omega_d), 
\end{equation}
at $\beta\neq{0}$,  considering the ordering of frequencies in Eq.(14) at $\lambda\neq{0}$.


Now we turn to our diagnosis. The exact expression for the {\it total} ($t$) energy shift is 
\begin{equation}
\Delta E_t(\omega_0,\omega_2)\, =\, [\Delta E (\Omega_0=\omega_0) + \Delta E(\Omega_0=\omega_2)],
\end{equation}
which is the sum of shifts in two independent modes. For two independent particles, constrained
by different ($j=d,e,w$) inputs from the exact informations, we obtain
\begin{equation}
\Delta E_t(\omega_j, \omega_j)\, =\, 2\, \Delta E(\Omega_0=\omega_j).
\end{equation}
Furthermore, at $\beta\rightarrow{\infty}$, from Eqs.(29-32) we can deduce 
\begin{equation}
\Delta E_t(\omega_0,\omega_2)\, =\, \Delta E_t(\omega_d,\omega_d),
\end{equation}
since $(2/\omega_d)=(1/\omega_0)+(1/\omega_2)$. However, this asymptotic agreement of a density-based, 
i.e., Kohn-Sham-like pre-optimization procedure with the exact 
total energy shift, which is an observable quantity, does not allow a similar
{\it prediction} for other quantities. 

For instance, the exact overlap 
$O=|<\Psi(x_1,x_2)|\Psi(x_1,x_2,t\rightarrow{\infty})>|^2$  becomes 
\begin{equation}
O=\frac{1}{\sqrt{1 +\Delta E(\omega_1)/\omega_1}}
\frac{1}{\sqrt{1 +\Delta E(\omega_2)/\omega_2}}=\sqrt{1-R(\omega_0)}\sqrt{1-R(\omega_2)} \nonumber
\end{equation}
This long-time form
differs (at $\lambda\neq{0}$) from the one obtained in the 
modeling with independent particles  where $O_d=|<\Psi_d(x_1,x_2)|
\Psi_d(x_1,x_2,t\rightarrow{\infty})>|^2=[1-R(\omega_d)]$. 
The above form for an overlap is an important result. It is expressed in terms of scattering
characteristics. From this point-of-view, it resembles to overlap-parametrizations
common in many-body physics with conventional {\it stationary} scattering theory.

Notice, that
in the very active research field of isolated confined cold-atom systems, with precisely specified
number of atoms, an overlap $O$ seems to be an experimentally  measurable quantity
\cite{Demler12,Islam15}. In that exciting field an abrupt ($a$) change \cite{Kagan96} at $t=0$, i.e., taking
the change $\Omega_0^2\Rightarrow{\Omega_f^2=(\Omega_0^2+\Lambda\omega_0^2)}>0$ abruptly from the initial 
to the final state, we have
\begin{equation}
R_a(\Omega_0,\Lambda)\, =\, \left[\frac{\Omega_0-\Omega_f}{\Omega_0+\Omega_f}\right]^2=
\left[\frac{1-\sqrt{1+\Lambda(\omega_0/\Omega_0)^2}}
{1+\sqrt{1+\Lambda(\omega_0/\Omega_0)^2}}\right]^2 \nonumber
\end{equation}
Experimentally, one can produce replicas via the controllable time-dependent tuning 
of confinement. In such a way, we would have a {\it series} of overlaps $O(\lambda,\Lambda)$. 
The analysis of these overlaps might allow to deduce 
statements on the interparticle interaction as well. Mathematically, a two-parametric
[$R(\omega_1)$ and $R(\omega_2)$] fitting could provide a more flexible framework than a
single-parametric [$R(\omega_d)$] one based on the mean-field concept.

\newpage

Next we calculate a quantum-mechanical Berry's connection \cite{Berry84,Simon83} which is
defined by $\mathcal{A}(t)\equiv{i\, <\Psi(X_1,X_2,t)|\partial_t\Psi(X_1,X_2,t)}>$, where $i=\sqrt{-1}$.
Considering the product-form of our $\Psi(X_1,X_2,t)$ in terms of two independent modes, 
we can proceed by using Eq.(18) together with Eq.(20). The one-mode ($1$) contribution 
to a two-term sum becomes
\begin{equation}
\mathcal{A}_1(t)= i<\phi(X,\Omega_0,t)|\partial_t\phi(X,\Omega_0,t)>= 
\frac{\Omega_0}{4}\left[\frac{\Omega^2(t)}{\Omega_0^2}B^2(t)+\frac{1}{B^2(t)}+
\frac{\dot{B}^2(t)}{\Omega_0^2}\right] \nonumber
\end{equation}
where, as we mentioned at Eq.(16) earlier, $\Omega^2(t)=[\Omega_0^2+\Lambda \omega_0^2 F(t)]$.
At the beginning we have $\mathcal{A}_1(t\rightarrow{-\infty})=\Omega_0/2$, since
at that limit $B(t)=1$ and $\dot{B}(t)=0$ due to $F(t)=0$. 

This constant $\Omega_0/2$ value will change during
the time-evolution and in the physically really important long-time limit, i.e., after perturbation,
we get
\begin{equation}
\mathcal{A}_1(t\rightarrow{\infty})= \frac{\Omega_0}{2}\, \frac{1+R(\Omega_0)}{1-R(\Omega_0)} 
\end{equation}
which is, as expected on physical grounds, the time-independent one-mode energy given in Eq.(24)].
One gets the original value $\Omega_0$ even at $t\rightarrow{\infty}$ iff $R=0$, i.e.,
in the discrete cases discussed at Eq.(29) for the attractive situation, i.e., with $\Lambda>0$. 

The mode contributions ($1$ and $2$) both will contain terms which oscillate during the action
of the external field. Therefore,
following the lead of an expert \cite{Simon83}, we think that it may be difficult to set up an
experiment to measure (at a given time) the phase picked up by an
evolving complex function when we solve an {\it initial-value} Cauchy problem using the
time-dependent Schr\"odinger equation. Due to that, the
associated action integral, on which a variational {\it boundary-value} problem in time can be based,
is generally nonstationary. This problem was discussed earlier \cite{Schirmer07} by focusing
on the foundation of TDDFT.

More generally, although our Cauchy problem rests on the time-dependent Schr\"odinger
equation which contains a first-derivative in time, the time-dependent exact solution for the
evolving normalized complex function requires two initial conditions to an underlying nonlinear
differential equation of Ermakov-type. The linearized, equivalent, version of such a nonlinear
equation would need three initial conditions. We believe that here is that mathematical complexity which
may prohibit to design, without {\it a priori} informations on initial conditions
or magnitudes of calculable observables, 
a completely predictive approximate method which is based on a Schr\"odinger-like 
auxiliary equation in orbital TDDFT.

\newpage

In Figures 1 and 2 we exhibit four-four curves which were 
calculated at $\lambda=3/8$ and $\omega_0=3$,
by taking $\Lambda={2/9}$ to Fig. 1, and $\Lambda=-2/9$ to Fig. 2. These
figures nicely illustrate our ordering in Eq.(31). The Hartree-Fock-based
curves (dotted) show the smallest energy shifts, in accord with
the over-localized character of the underlying wave function.
The density-based Kohn-Sham-like approximation, dash-dotted curves,
gives the closest values to the exact results (solid curves). 
There is a {\it crossing} \cite{Nagy11} point between these curves, at around $\beta=2.86$, 
independently of the sign of $\Lambda$. But, despite of the under-localized
character of the density-optimal independent-particle approximation, it approximates
reasonably well the exact two-mode results (solid curves) which are based on different frequencies.

This, {\it a posteriori} observation supports the folklore based on the general practice with auxiliary
orbitals in the mean-field TDDFT. Unfortunately, as the crossing found here signals, one can not be
sure {\it a priori} about a prediction on the from-below or from-above character of numerical results.
Furthermore, a comparison of magnitudes of the corresponding curves in Figs. 1-2 shows that the
deviation from the exact results (solid curves) can be larger in the case of repulsion
where $\Lambda<0$. This in in harmony with our expectation, based on the 
early results of esteemed experts \cite{Fermi40,Teller47}. Indeed, close to 'ionization' a precise
description is needed in order to characterize the magnitude of an observable energy shift.

\begin{figure}
\scalebox{0.3}[0.3] {\includegraphics{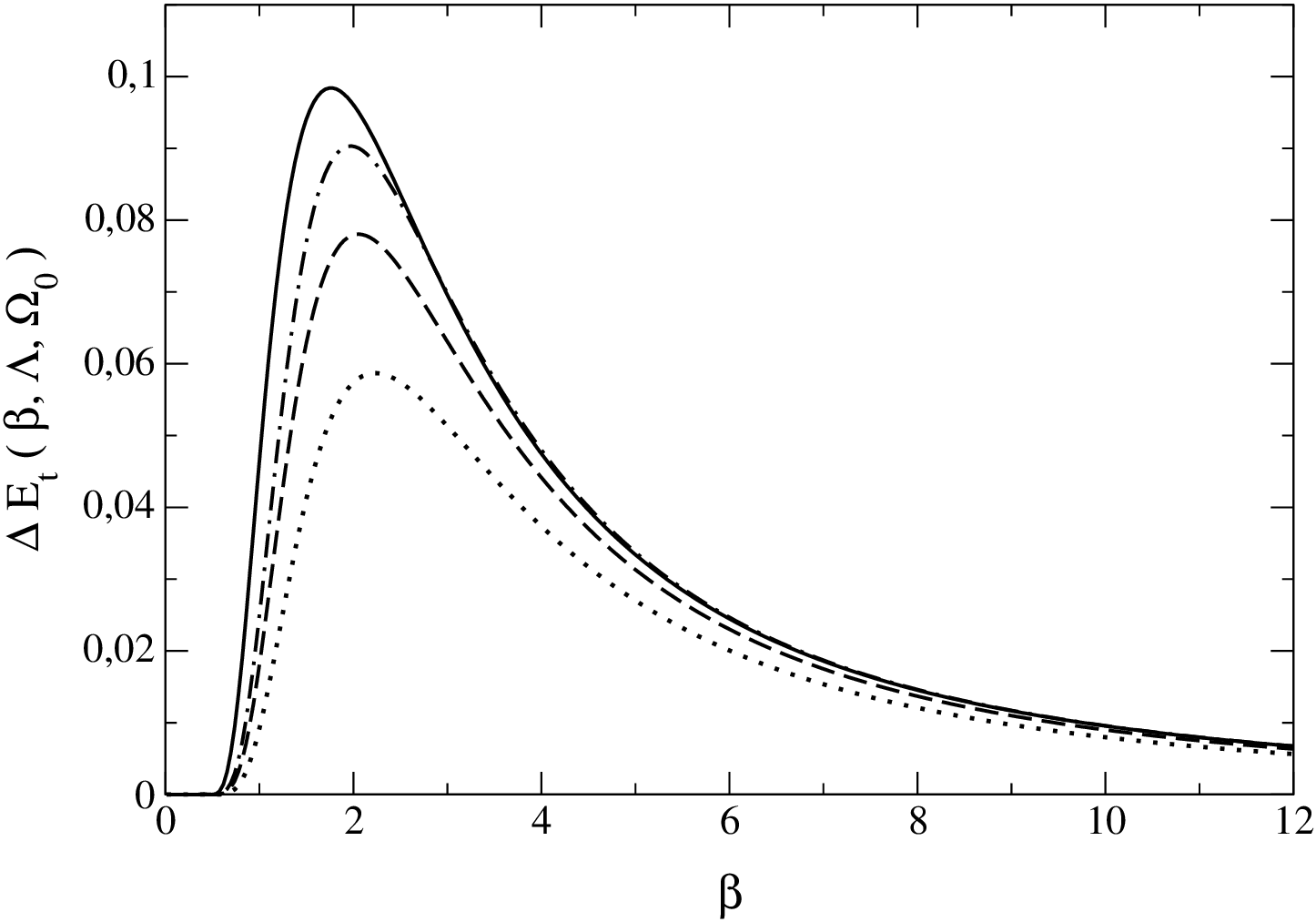}} \caption
{Total energy shifts, denoted by $\Delta E_t(\beta,\Lambda,\Omega_0)$,
in the attracive ($\Lambda=2/9$) case, as a function of $\beta$ at $\lambda=3/8$. Thus,
the input frequencies are
$\omega_0=3$, $\omega_2=1.5$, $\omega_e=2.372$, $\omega_w=2.121$, and $\omega_d=2$
to Eq. (29)  and Eqs.(32-33). Solid curve: exact result. Dotted curve: Hartree-Fock-like
approximation. Dash-dotted curve: density-optimal, Kohn-Sham-like approximation.
Dashed curve: approximation based on the first natural orbital with unit occupancy. 
These approximations are based on pre-optimized initial states in Eq.(4), Eq.(7), and Eq.(12), respectively.
\label{figure1}}
\end{figure}

\begin{figure}
\scalebox{0.3}[0.3] {\includegraphics{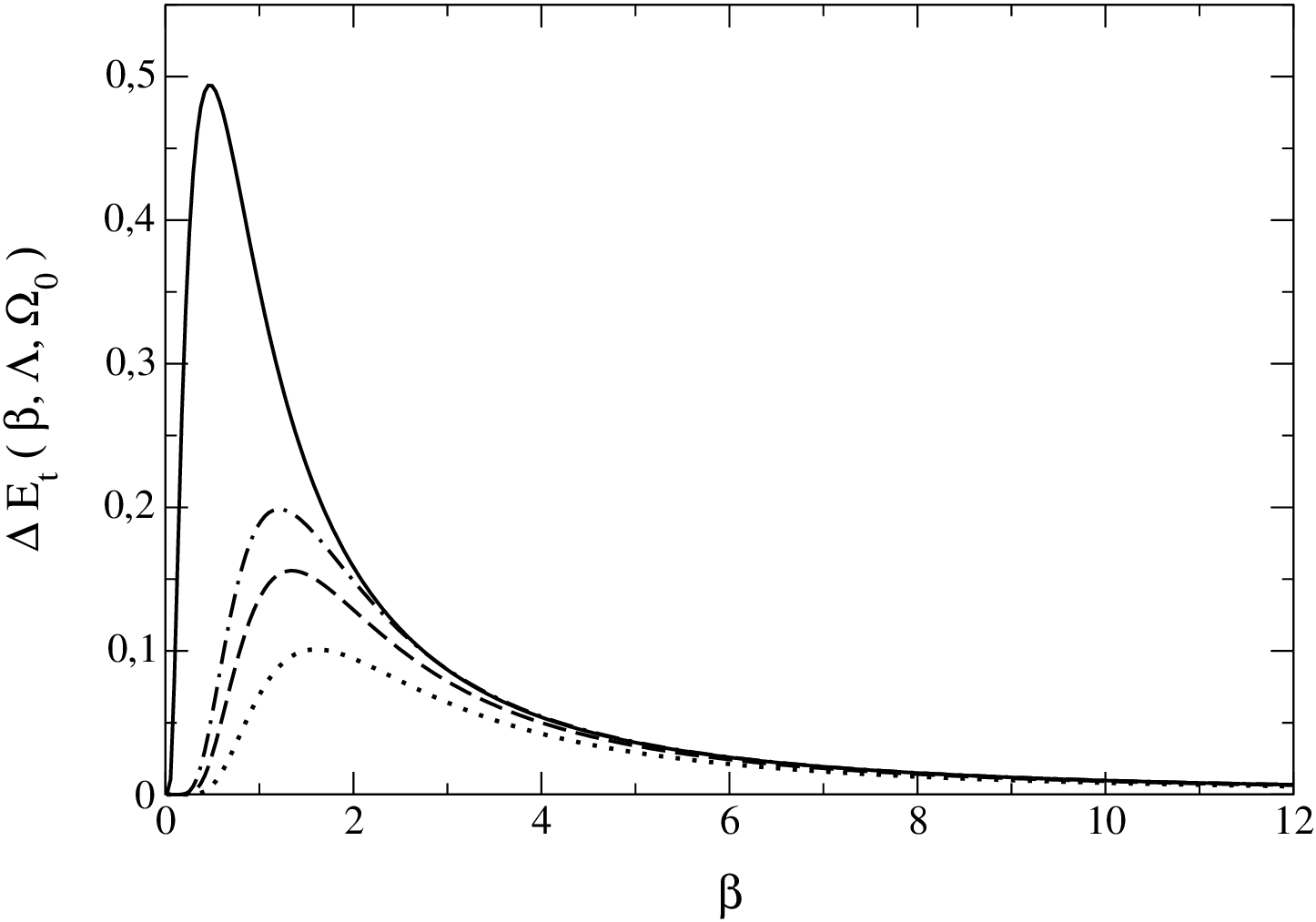}} \caption
{Total energy shifts, denoted by $\Delta E_t(\beta,\Lambda,\Omega_0)$,
in the repulsive ($\Lambda=-2/9$) case, as a function of $\beta$ at $\lambda=3/8$. Thus, the
input frequencies are
$\omega_0=3$, $\omega_2=1.5$, $\omega_e=2.372$, $\omega_w=2.121$, and $\omega_d=2$
to Eq. (29)  and Eqs.(32-33). Solid curve: exact result. Dotted curve: Hartree-Fock-like
approximation. Dash-dotted curve: density-optimal, Kohn-Sham-like approximation.
Dashed curve: approximation based on the first natural orbital with unit occupancy. 
These approximations are based on pre-optimized initial states in Eq.(4), Eq.(7), and Eq.(12), respectively.
\label{figure2}}
\end{figure}

\eject

We mentioned at Eq.(29) that the $\beta\propto{v}$ choice ($v$ is the velocity of 
a fast charged intruder) seems to be
a reasonable one in the sudden limit \cite{Artacho07}. As a preliminary step to
future realistic calculations, here we estimate a characteristic interaction-time
in classical atom-atom collision with reduced mass $\mu=M_1M_2/(M_1+M_2)$. We model
\cite{Nagy94,Arista04} the screened atom-atom interaction via a finite-range, $r\in{[0,r_0]}$,
Mensing-type potential
\begin{equation}
V(r)\, =\, \frac{Z_1\, Z_2}{r}\left(1 - \frac{r}{r_0}\right). \nonumber
\end{equation}
and solve the classical problem for the finite collision time ($T_c$) by integration.
We get 
\begin{equation}
T_c(b,v,r_0)=\frac{2}{v}\left[\frac{\sqrt{r_0^2-b^2}}{p}+
\frac{Z_1Z_2}{2E}\frac{1}{p^{3/2}}\ln\frac{\sqrt{p}
\sqrt{r_0^2-b^2}+r_0+(Z_1Z_2/2E)}{\sqrt{(Z_1Z_2/2E)^2+p\,
b^2}}\right] \nonumber
\end{equation}
where $E=\mu v^{2}/2=M_2/(M_1+M_2)E_1$ is the kinetic energy. The heavy charged projectile ($Z_1,M_1$)
collides at impact
parameter $b$ and energy $E_1=M_1 v^2/2$ with a target atom ($Z_2,M_2$) at rest.
The dimensionless parameter introduced is $p=[1+(Z_1Z_2/r_0)/E]$,
i.e., it is the ratio of potential and kinetic energies.
A closer inspection shows that one may apply
\begin{equation}
T_c(b,v,r_0)\, \simeq{\alpha(v)\, \frac{r_0}{v}\,
\frac{E}{E+Z_1Z_2/r_0}}\, \sqrt{1-\frac{b^2}{r_0^2}} \nonumber
\end{equation}
to get a reasonable estimation as a function of the heavy
projectile velocity $v$. The $\alpha(v)$ parameter decreases
from 4 (at $v\rightarrow{0}$) to 2 (at $v\rightarrow{\infty}$). A cross-sectional average
becomes 
\begin{equation}
\bar{T}_c(v,r_0)=\alpha(v)\, \frac{r_0}{3v}\, \frac{E}{E+Z_1Z_2/r_0}. \nonumber
\end{equation}
In this simple modeling of the nuclear ($Z_1Z_2>0$)
elastic collision we are at the sudden situation ($\bar{T}_c\rightarrow{0}$) for {\it both}
limits of the intruder velocity $v$.

In particular, our asymptotic form in Eq.(30) would give, with the natural choice of $2\beta=(1/\bar{T}_c)$,
a quadratic behavior $\Delta E(\Omega_0)\propto{v^2}$ at small enough velocities. Such an
energy shift can be much smaller than the energy difference ($\sim{\Omega_0}$) between the
unperturbed states. However, a time-dependent excitation process is a rigorously quantum mechanical
process since, classically, this energy shift is not enough to make a single-particle jump 
for which one could take conservation laws of two-body kinematics as the only decisive constraints.

In Figure 3, we plot the ratio $\{[\Delta E_t(\Lambda=-2/9)/\Delta E_t(\Lambda=2/9)] -1\}$
as a function of the velocity for which we take for simplicity $v=\beta$.
This corresponds to a quasiclassical Thomas-Fermi-like choice 
$r_0=3/4\simeq{1/Z_2^{1/3}}$ (with $Z_2=2$) to be used in the above high-velocity form. 
The corresponding results are plotted for $v\in[4,12]$.
One can see that the plotted difference approaches zero, as expected on physical grounds, by
increasing the velocity. At the lower range (still swift antiprotons or protons, with
bombarding energy about $E_1=400-500$ keV) the difference can be in the $10\%$ range. This
magnitude is not in contradiction with the estimation ($15\pm{5}\%$, at that energies) based on
OBELIX stopping data obtained \cite{Rizzini04} at CERN for a gaseous He target ($Z_2=2$)
with $Z_1=\pm{1}$.

\begin{figure}
\scalebox{0.3}[0.3] {\includegraphics{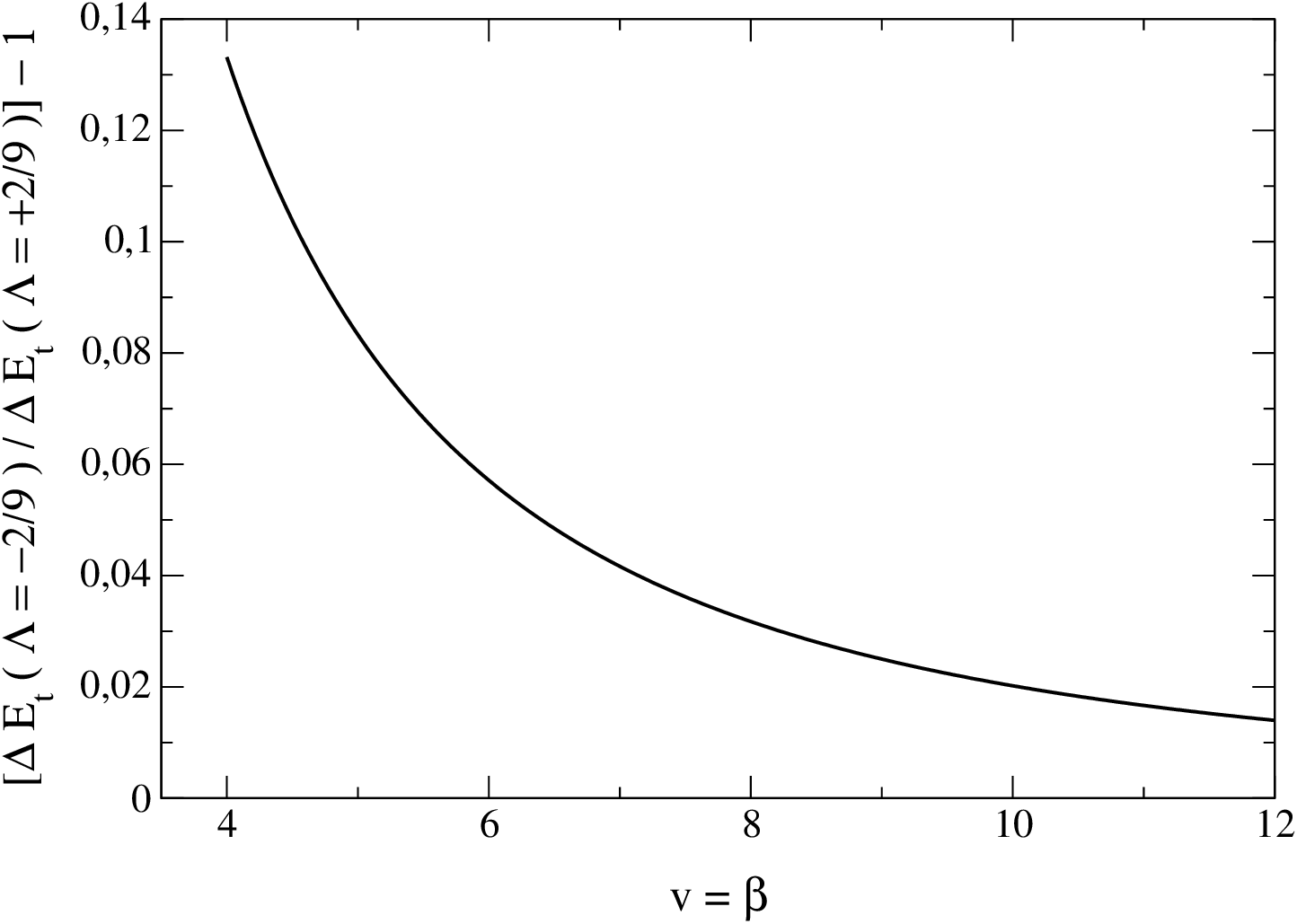}} \caption
{The exact ratio in form of $\{[\Delta E_t(\Lambda=-2/9)/\Delta E_t(\Lambda=2/9)] -1\}$
as a function of the velocity $v$ for which we take $v=\beta$. 
In our parametrization we have $(\Lambda\omega_0^2/2)={\pm{1}}$.
\label{figure3}}
\end{figure}

We left a more detailed investigation of the energy shift, by using an impact parameter dependent estimation
for the collisional time $T_c(b,v,r_0)$, for a future study. Such a study, combined with estimations
for the distant (large impact parameter) range, where the polarization-related energy shift 
generated by a dipole-like field may be important \cite{Kabachnik90}, 
is highly desirable for an {\it interacting} system. Furthermore, one should 
keep in mind that contributions from
the nuclear-stopping and electronic-stopping are evidently correlated \cite{Sigmund08}.
In our case, a simple re-parametrization with $r_0(v)>r_0$ would change the original
partial stopping calculated at $r_0$. At high velocities both, nuclear and electronic,
contributions could be larger due to their scaling $\propto{[r_0(v)]^2}$.
Consistent attempts \cite{Correa12} are desirable.


\section{Summary}

In this contribution we determined the exact solution for the ground-state wave function of a
two-particle correlated model atom with harmonic interactions. The model was introduced by
Heisenberg and constitutes a cornerstone in such areas of physics, as the field
of correlated atoms and confined quantum matter.
Next, a time-dependent external field of finite duration is addded to the Hamiltonian and the
resulting Cauchy problem is solved. The impact of this external field is investigated by calculating
the energy shift in the model atom. It is found that
the nonperturbative energy shift reflects the sign of the driving field. 
Three independent-particle approximations, defined from exact informations on the correlated model, 
are investigated as well in order to understand their limitations.

Paralell to the investigation of the sign-effect in an energy shift, we determined and analyzed other
{\it measures} of an inseparable two-body problem, namely the overlap and a Berry's
connection which are also based on the precise wave function. To our best knowledge,
the overlap seems to be a measurable quantity in modern experiments with confined
cold atoms of controllable number. Unfortunately we can not see, at this moment at least, an
experimental tool to measure Berry's connection in our time-dependent problem. 

But, to be optimistic at this alarming point as well, we would like to refer to
the wish of the Duke of Gloucester in one of Shakespeare's famous plays. 
{\it I can add colours to the chameleon} -- emphasized the Duke in his monologue. We 
believe that by the application of carefully selected
experimental and theoretical tools, finally one can get a transferable knowledge on
an observable 'chameleon' by focusing on its really visible 'colours'.

\begin{acknowledgments} This theoretical work is dedicated to the memory 
of Professor Rufus Ritchie for his outstanding contribution to the theory of the stopping power.
The kind help of
Professor Nikolay Kabachnik is gratefully acknowledged. This work was supported partly
by the spanish Ministry of Economy and Competitiveness (MINECO: Project FIS2016-76617-P).
\end{acknowledgments}
%


\end{document}